\documentclass[useAMS,usenatbib]{mn2e}
\def\arcdeg{\hbox{$^\circ$}}
\def\arcmin{\hbox{$^\prime$}}

\def\fm{\hbox{$.\!\!^{\rm m}$}}

\def\fdg{\hbox{$.\!\!^\circ$}}

\def\farcs{\hbox{$.\!\!^{\prime\prime}$}}
\def\agt{\mathrel{\hbox{\rlap{\hbox{\lower4pt\hbox{$\sim$}}}\hbox{$>$}}}}

\usepackage{epsfig}

\title[The open cluster Trumpler 20]{Trumpler 20 - an old and rich open cluster
\thanks{Based on observations collected at La Silla Observatory, ESO
(Chile) with the FEROS spectrograph at the MPG/ESO 2.2-m telescope (program
ID 080.D-2002(A)).
}
\thanks{Table 1 is available in electronic form at the
CDS via anonymous ftp to cdsarc.u-strasbg.fr (130.79.128.5)
or via http://cdsweb.u-strasbg.fr/cgi-bin/qcat?J/MNRAS/ }
\thanks{This is paper 35 of the WIYN Open Cluster Study (WOCS). }
}
\author[I. Platais et al.]
{I. Platais,$^{1}$
\thanks{E-mail: imants@pha.jhu.edu (IP); cmelo@eso.org (CM);
jfulb@pha.jhu.edu (JPF); verap@stsci.edu (VKP); 
pedro.figueira@obs.unige.ch (PF); barnes@lowell.edu (SAB);
rmendez@das.uchile.cl (RAM)
}
 C. Melo,$^{2}$  J.P. Fulbright,$^{1}$ V. Kozhurina-Platais,$^{3}$ P. Figueira,$^{4}$ 
\newauthor  S. A. Barnes,$^{5}$ R. A. M\'{e}ndez$^{6}$\\
$^{1}$Department of Physics \& Astronomy, The Johns Hopkins University,
Baltimore, MD 21218, USA\\
$^{2}$European Southern Observatory, Karl-Schwarzschild-Str. 2, Garching
bei M\"{u}nchen, Germany\\
$^{3}$Space Telescope Science Institute, 3700 San Martin Drive, Baltimore,
MD 21218, USA\\
$^{4}$Observatoire de Gen\`{e}ve, 1290 Sauverny, Switzerland\\
$^{5}$Lowell Observatory, Flagstaff, AZ 86001, USA\\
$^{6}$Departamento de Astronomia, Universidad de Chile, Casilla 36-D, Santiago,
Chile
}
\begin{document}

\date{Accepted 2008 September 25. Received 2008 September 25; in original form 2008 August 7}

\pagerange{\pageref{firstpage}--\pageref{lastpage}} \pubyear{2008}

\maketitle

\label{firstpage}

\begin{abstract}
We show that the open cluster Trumpler~20, contrary to the earlier findings,
is actually an old Galactic open cluster. New CCD photometry and
high-resolution spectroscopy are used to derive the main parameters of
this cluster. At [Fe/H]=$-$0.11 for a single red giant star, the metallicity
is slightly subsolar. The best fit to the color-magnitude diagrams
is achieved using a 1.3 Gyr isochrone with convective overshoot.
 The cluster appears to have
a significant reddening at E(B-V)=0.46 (for B0 spectral type), although
for red giants this high reddening yields the color temperature exceeding the
spectroscopic $T_{\rm eff}$ by about 200~K. Trumpler 20 is a very rich
open cluster, containing at least 700 members brighter than $M_V=+4$.
It may extend over the field-of-view available in our study at 
20$\arcmin\times$20$\arcmin$.
\end{abstract}

\begin{keywords}
Open clusters and associations: individual: Trumpler 20 -- stars:abundances.
\end{keywords}

\section{Introduction}

The number of old open clusters (age $\agt$1 Gyr) is $\sim$60
\citep{fr95} or only $\sim$4\% of the all known clusters in the Milky Way
\citep{di04}. Because of this small number of such clusters and the fact that
many of them are poorly populated, any addition to this body of old
 clusters is highly desirable.
Here we report a serendipitous discovery of an old open cluster among
the known registered clusters in
the WEBDA\footnote{http://www.univie.ac.at/webda/}. It seems to be
enlightning and worthy to reflect on what initially led us to
this object. In March of 2008 we had an observing run at the 2.2m
MPG/ESO telescope with the FEROS spectrograph. For the morning hours,
there were no regular targets available and, hence, we started to look
for some old open cluster with 12$^{\rm h}$$<$RA$<$16$^{\rm h}$.
There are 7 such clusters in the WEBDA but none of them appeared to be
well-suited for our instrumental setup or offering a compelling science.
A search for other potential targets led us
to a large photometric survey of 55 young open clusters in the southern
sky by \citet{mc05}. The primary goal of their survey was to examine the age
and evolutionary dependence of the Be star phenomenon.
Examining the provided color-magnitude diagrams (CMD)
our attention was caught by the CMD of Trumpler~20. It was obvious that for
this cluster the authors had missed out a characteristic signature of all
old open cluster -- that is the red clump, prominently displayed in its
CMD at $y=14.5$ and $b-y=0.9$ \citep[Fig. 59]{mc05}. These authors
opted  for a young 160 Myr cluster, instead.  The apparent lack of reliable
parameters for Trumpler~20 (Tr~20 hereafter) prompted the present study.

In the discovery paper \citep{tr30}, Tr~20 (=C1236-603) is
described as a "rich cluster of
regular outline composed exclusively of very faint stars nearly evenly
scattered" and classified as a type III~2r cluster. 
Using the calibrations between the estimated diameter
in parsecs and the apparent angular diameter, for this cluster with an angular
diameter of 10$\arcmin$ Trumpler reports a distance equal to 2240~pc.
A few decades later, \citet{ho65} estimated the total
number of cluster members from starcounts to be at $\sim$200 down to
$m_{pg}$$\sim$17.  The Str\"{o}mgren $by$ photometry given in \citet{mc05} has
a limited potential in deriving a photometric distance of Tr~20 because
it covers as little as $\sim$1.5 mag of its main sequence. One helpful
hint from the paper by \citet{mc05} is the noted relatively high reddening
in this area of the sky at $E(B-V)$$\sim$0.35.  For the open clusters
located within $2\degr$ from Tr~20, the WEBDA database
offers additional reddening estimates, reaching up to $E(B-V)=0.4$.

The access to high-resolution spectroscopy at the 2.2m MPG/ESO gave us
an opportunity to measure radial velocities for a few possible Tr~20
members and even attempt to measure its metallicity [Fe/H]. Another
fortunate concurrence involved an access to photometric  observations
at the 1m telescope of the Cerro Tololo Inter-American Observatory (CTIO).
These two facilities allowed us to collect the necessary data and were
instrumental in obtaining fundamental properties of Tr~20.

\section{CCD $BVI$ photometry}

In 2008 March 20-22, new photometric CCD observations of Tr~20 were obtained
at the CTIO 1-m telescope, operated by the Small and Moderate Aperture Research Telescope System (SMARTS) consortium. The instrumental setup includes the
Y4K Cam consisting of an STA 4064$\times$4064 CCD with 15-$\umu$m pixels
mounted in the LN$_{2}$ dewar. The pixel scale is $0\farcs289$ pix$^{-1}$,
allowing to cover a field-of-view of 20$\arcmin\times$20$\arcmin$.

\begin{table*}
\begin{minipage}{140mm}
  \caption{$BVI$ photometry in Tr~20 (a sample only).}
  \begin{tabular}{@{}rcccrrccccc}
  \hline
   ID  & $\alpha$$_{\rm J2000}$ & $\delta$$_{\rm J2000}$ & $V$ & $B-V$ &
$V-I$ & $\epsilon_V$ & $\epsilon_{B-V}$  & $\epsilon_{V-I}$ & $n_{BV}$ & $n_{VI}$ \\
 \hline
   1 & 189.67025 & $-$60.80123 & 12.746 & 0.686 & 1.030 & 0.010 &  0.014 & 0.014 & 2& 2\\
   2 & 190.04024 & $-$60.80076 & 16.684 &  1.117 & 1.638 & 0.010 & 0.031 & 0.019 & 1& 2\\
   3 & 189.96685 & $-$60.79839 & 16.405 &  0.804 & 1.058 & 0.013 & 0.028 & 0.019 & 2& 2\\
   4 & 190.08571 & $-$60.79829 & 17.084 & 99.999 & 2.112 & 0.017 & 0.000 & 0.022 & 0& 2\\
   5 & 190.12292 & $-$60.79595 & 15.238 &  0.738 & 1.042 & 0.011 & 0.018 & 0.016 & 2& 2\\
\hline
\end{tabular}
\end{minipage}
\end{table*}

Our nightly observations in the Johnson-Cousins $BVI_{\rm C}$ filters
 consisted of 10 domeflats
in each filter, imaging of three \citet{la92} standards (PG1323, PG1633,
Mark~A), and imaging the cluster itself using its nominal center from WEBDA.
We recognize that the color range of our Landolt standards is
relatively narrow, ranging from $-$0.24 to $+1.13$ in $B-V$
(only one star at $B-V=1.49$). This may affect the accuracy of
color terms in transformations to the standard system, especially
for red stars. 
Exposure times for Tr~20 ranged from 10 to 600~s, depending on a filter, 
so that in each filter there was a short and a longer exposure. This
arrangement was kept in each of these three nights. One potential drawback
of our observations is a relatively high airmass (1.49-1.74) at which
the cluster was observed. It degraded the seeing down to $\sim$$1\farcs7$. 

Although the Y4K Cam is a single-chip CCD imager, it has four
sections (quadrants) operated by independent amplifiers. Therefore,
each section has its own bias level and gain. For initial reductions
(triming, bias, flat-field and illumination corrections)
we used the IRAF scripts
prepared by P.~Massey\footnote{http://www.lowell.edu/users/massey/obins/y4kcamred.html} specifically for this camera. We note that all raw CCD images contain
a number of "dust ghosts" which, unfortunately, cannot be completely
eliminated. In most severe cases the residual ghosts can be on the order of
a few per cent of the background level.

We used the IRAF routines DAOPHOT and PHOT to perform the aperture photometry
on our CCD images. To generate the lists of instrumental magnitudes,
an in-house IRAF script was applied to all images. The aperture correction
was derived from instrumental magnitudes through 8- and 12-pixel apertures.
Our magnitudes are derived using an 8-pixel aperture. Conventional calibration
equations (e.g., \citealt{ca08}) were used to derive photometry
in the standard $BVI_{\rm C}$.
Our calibration coefficients are fairly close to those listed by
\citet{ca08} for the 2005 season with the exception of a sign
for all three color terms, which, we believe, is incorrect in their study.

The catalog of CCD photometry is constructed by combining all calibrated
magnitudes in each of the three filters. A photometric measurement is not
included if its formal error exceeds 0.08 mag, effectively enforcing a cut
at $V$$\sim$18.5 mag. In order to qualify for
an entry in the catalog, a star must have at least three detections
in $V$ filter. The mean magnitudes are calculated by weighted 
individual measurements by the formal photometric errors.

The astrometric calibration of pixel coordinates was performed using the 
UCAC2 stars \citep{za04} as a reference frame. The coordinate transformation
required a plate model with with the linear and quadratic terms, including
the two main cubic distortion terms. At the epoch of 2008.22, the average
standard error of a solution is $\sim$75~mas. The catalog (complete Table~1
is available in electronic form only) contains identifier, 
J2000 equatorial coordinates, $BVI$ photometry and its formal errors
for 7166 stars. Last two columns indicate the number of image pairs
available for each of the $BV$ and $VI$ indices.

The only external source of CCD photometry in a bandpass reasonably
close to the Johnson $V$ is that of \citet{mc05}. 
The differences between our $V$ magnitudes and Str\"{o}mgren $y$
can be approximated by a linear relationship $V-y=0.008-0.035(B-V)$, showing
the residual scatter (standard error) of 0.022 mag. The slope in this
relationship has an opposite sign, if compared to a similar transformation
between the Johnson and Str\"{o}mgren visual magnitudes for unreddened stars:
$V-y=+0.038(B-V)$ \citep{co85}.  Such a sign flip is clearly
related to the effect of a considerable reddening of stars in the area
of Tr~20. Similarly, for the color differences, we obtained the
following linear regression: $(B-V)-(b-y)=-0.092+0.373(B-V)$ with a standard
error of 0.030. A direct transformation from Str\"{o}mgren $(b-y)$ to $(B-V)$
yields: $(B-V)=-0.085+1.497(b-y)$ and a standard error of 0.057 mag. This
transformation cannot directly be compared to the published
relationships because an important term which includes the Str\"{o}mgren
index $m_1$ is missing. Overall, our $BV$ photometry appears to be on
the standard $UBV$ system.

\section{Spectroscopy with FEROS}

The spectroscopic observations of Tr~20 at the MPG/ESO 2.2m telescope in La
Silla Chile, using the two-fiber FEROS echelle spectrograph
\citep{ka99}, were collected exactly on the same dates as
our photometry. The high resolution of this spectrograph
($R\sim50000$) and its wide wavelength range (360-920~nm) makes this
instrument very useful in the chemical abundance studies, allowing us to
select unblended weak iron lines. The observations were obtained in the
Object-Sky mode in which one fiber collects the light from target
and the other fiber -- from the sky background.

We used \citet{mc05} photometry to select a handful of spectroscopic targets
for Tr~20. These included three bright stars from the tip of the red giant
branch (RGB) and three fainter stars from the red clump (Table~2). While
these stars are still bright enough for radial velocity determination
with the FEROS setup, a high enough signal-to-noise (S/N) ratio required
for abundance analysis is problematic for a $\sim$13.5 magnitude star.
Hence, for a chosen chemical abundance star
3764 = MG~675, we obtained four separate exposures spread over all three
nights and totalling three hours.

\subsection{Radial velocities}

The reductions of all spectra were performed using the standard FEROS
pipeline, which yields a 1-D re-binned spectrum sampled at 0.03\AA~ steps.
The radial velocities have been derived using the cross-correlation techniques
and a K0~III spectral type digital mask as the template (\citealt{ba96,qu95}).
A moderate S/N of our spectra at 30-50 yields the
radial velocity photon noise errors on the order of 25~m~s$^{-1}$.
The final uncertainty of radial velocity is limited by the overnight
spectrograph drift due to the changes in the index of air refractivity and
atmospheric pressure. A couple of standard stars were used to account for
this drift, described in detail by  \citet{pl07}. The formal
precision of this correction is about 20~m~s$^{-1}$. 
Thus, the estimated accuracy of our radial velocity determination
is $\sim$30~m~s$^{-1}$ per single observation.
The repeat observations
of 3764 = MG~675 indicate the standard error of its radial velocity to
be at the level of 25~m~s$^{-1}$. The cluster's mean heliocentric radial
velocity is estimated to be $-40.8$~km~s$^{-1}$, excluding from the sample
star 4645. The mean heliocentric radial velocities are given in Table~2.
First column is our star number; second -- star number from \citet{mc05}.
Remaining columns are self-explaining. Only two stars (3764, 4645)
are observed more than once.  The last 7 stars are tentative
photometric sub-giant branch stars in Tr~20 for which we do not have radial
velocities. It is a high priority to establish their cluster membership
via radial velocities because these stars are crucial in anchoring
the isochrone fit (see Sect.~4). 

\begin{table}
  \caption{Heliocentric radial velocities in Tr~20.}
  \begin{tabular}{@{}crccc}
  \hline
   ID  & MG & $V$ & $B-V$ & $<$RV$>$~km~s$^{-1}$ \\
 \hline
3309 &  794 & 13.787 & 1.524 & $-40.85$ \\
3317 &  791 & 14.590 & 1.306 & $-39.43$ \\
3730 &  685 & 14.579 & 1.278 & $-42.20$ \\
3764 &  675 & 13.625 & 1.521 & $-40.86$ \\
4645 &  454 & 13.680 & 1.501 & $-44.44$ \\
4684 &  442 & 14.492 & 1.279 & $-40.76$ \\
\hline
1441 &      & 15.605 & 1.074 &  \\
2156 & 1055 & 15.876 & 1.201 &  \\
3606 &  715 & 15.844 & 1.155 &  \\
3847 &  654 & 15.777 & 1.155 &  \\
4498 &      & 15.741 & 1.222 &  \\
6117 &  132 & 15.674 & 1.116 &  \\
6241 &  116 & 15.855 & 1.191 &  \\
\hline
\end{tabular}
\end{table}

\subsection{[Fe/H] determination}

Limited by the requirement to reach an adequate S/N, we focussed our
attention just on star 3764 = MG~675. Its radial velocity appears to
be stable and is very close to the mean velocity of Tr~20, thus
minimizing the chance that the star might be a binary. We consider it
a typical representative of the entire cluster.  The final spectrum
 is combined from four individual
exposures that total 10778~s of the open shutter time.  The data were extracted
using a standard pipeline routine, and the final spectrum
has a signal-to-noise ratio of $\sim$65 per resolution element near 6100~\AA.
 
The parameter determination of 3764 = MG~675 was based on the analysis of
\citet{fu06}.  We have adopted the list of iron lines from \citet{fu06}.
The lines in this
list were selected to be those least affected by blending for use in the
study of metal-rich bulge giants, where the differential analysis was
done relative to Arcturus.
In \citet{fu06}, the zero point of the differential metallicity scale was
set by another differential analysis between the Sun and Arcturus.  They
adopted a solar iron abundance of log$\epsilon$(Fe)=7.45. The lines
were measured manually using the IRAF {\it splot} package.
We do not use lines stronger than $\sim$120~m\AA.

The effective temperature ($T_{\rm eff}$), surface gravity, and microturbulent
velocity ($v_{\rm t}$) of 3764 = MG~675 were set by
the excitation temperature method as described in \citet{fu06}.  If we
adopt a mass of 1.5~M$_\odot$ and $M_{\rm V}$=$-$0.2, we determine
$T_{\rm eff}$=4321$\pm$59~K, log~g=1.82, and
$v_{\rm t}$=1.80$\pm$0.08~km~s$^{-1}$.  For the 107 Fe~I lines analyzed,
we find a mean abundance of log$\epsilon$(Fe)=7.34$\pm$0.13 and
for the 5 Fe~II lines we find 7.48$\pm$0.09. Therefore, we adopt
[Fe/H]=$-$0.11$\pm$0.13 for the entire Tr~20.
The difference in the iron abundances as determined by the
neutral and ionized lines would be lessened if the mass of the star were
slightly lower or the adopted absolute magnitude were slightly brighter.

\citet{mc05} give $(b-y)$=1.046 for 3764 = MG~675.  If we adopt
$E(B-V)$=0.46 (see Sect 4.1)
and $E(b-y)$=0.745$E(B-V)$, we find $(b-y)$$_0$=0.703.  The
color-$T_{\rm eff}$ relations for giant stars by \citet{ra05}
yield a photometric $T_{\rm eff}$ of 4500~K for this star at [Fe/H]=$-0.11$~dex.
Our spectroscopic $T_{\rm eff}$ is 180~K cooler, which corresponds to the
star being $\sim$0.06~mag bluer in $(b-y)_0$, possibly over-corrected for
the reddening.

\section{Color-magnitude diagram and isochrone fit}

As the result of this study, new CMDs in $BV$ and $VI$ can be constructed
for analysis (see Table~1). 
Although the nominal depth of deep $V$ frames is $V$$\sim$20.5,
the considerably shallower $B$ and $I$ frames limit our CMDs to 
$V$$\sim$18-18.5. We do not reproduce here the observational
CMDs. Instead, our goal is to obtain reddening-free CMDs which then
can be fitted with the theoretical isochrones. Such color-magnitude
diagrams have been constructed only for stars located inside
the circle with a radius of 8$\arcmin$, centered onto the cluster. 
Owing to the lack of kinematic membership, the chosen size of field
appears to be a reasonable compromise between a desire to include the
majority of cluster members and minimizing the impact of field star
contamination. 

\begin{figure}
\includegraphics[width=84mm]{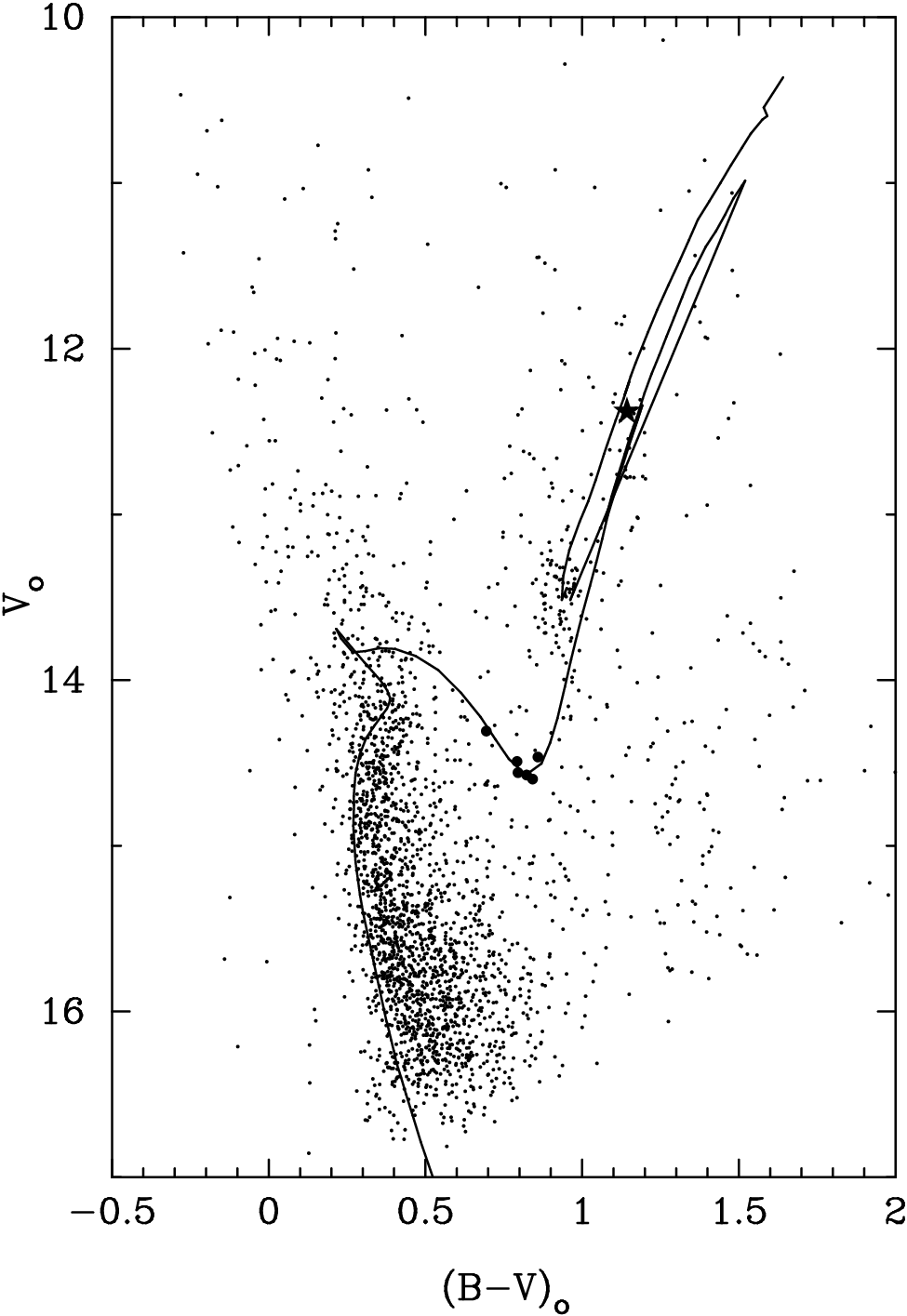}
 \caption{A fit of 1.3-Gyr isochrone to the $BV$ color-magnitude diagram (CMD)
of Tr~20. The bold points indicate the tentative sub-giant branch stars
(see Table~2). Star 3764 is marked by the asterisk. 
The two main cluster parameters
-- reddening and distance modulus -- applied to the observational CMD
and the fit itself are listed in Table~3. Note that here and in Fig.~2
we plot only those stars which have complete $BVI$ photometry. There are
2452 such stars.
}
\end{figure}

\begin{figure}
\includegraphics[width=84mm]{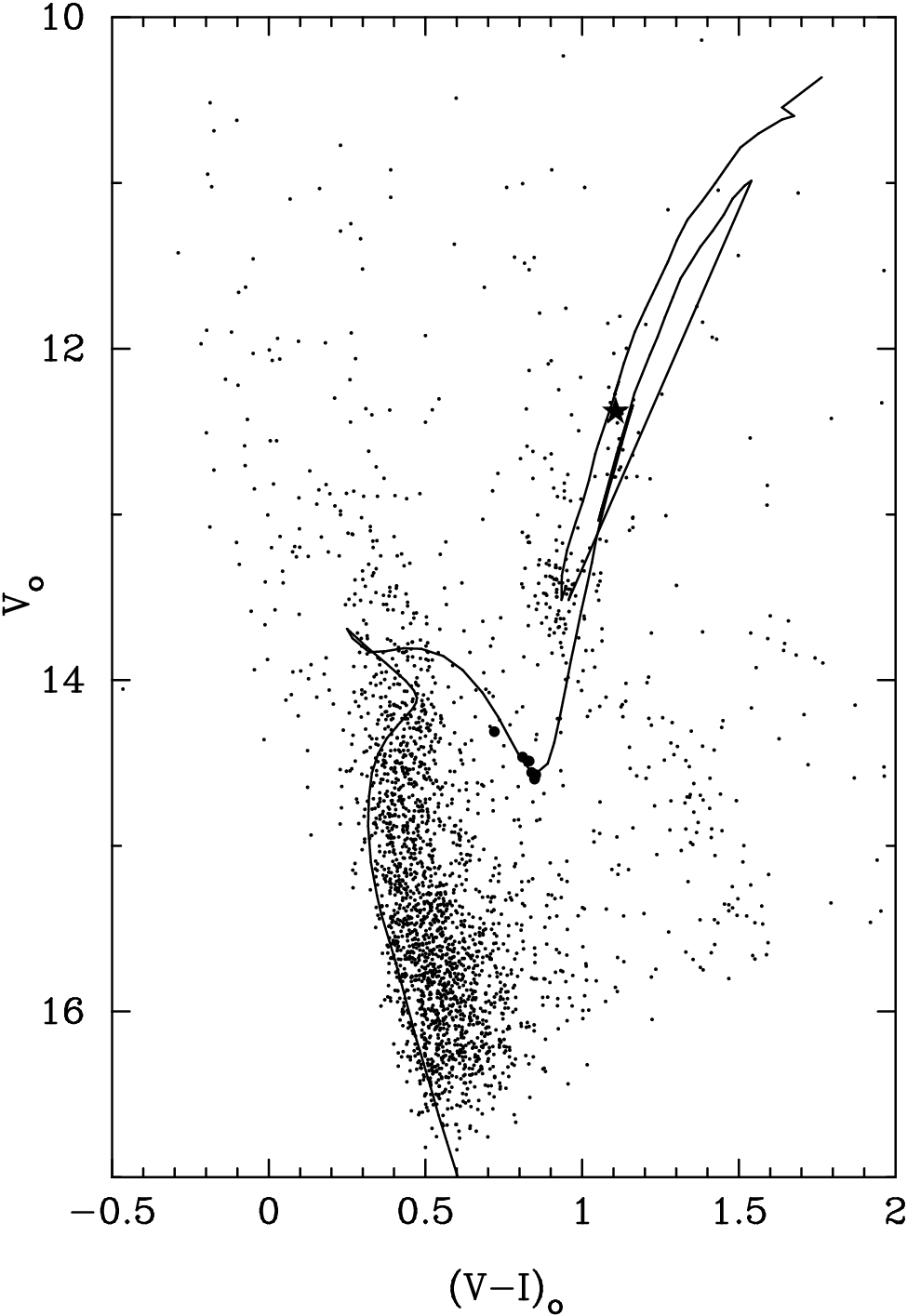}
 \caption{A fit of 1.3-Gyr isochrone to the $VI$ CMD of Tr~20.
The bold points indicate the tentative sub-giant branch stars.
The reddening adjusted to $V-I$ color index and distance modulus are the
same as in Fig.~1.
}
\end{figure}

\subsection{Interstellar reddening and age}

For evolved F-G spectral class main sequence stars, the determination of
reddening is not trivial. This point is well-illustrated by critical analysis
of all published $E(B-V)$ values for M67 \citep{ta07}, ranging
from $-0.01$ to $+0.14$! Our analysis of reddening is further complicated
by the fact that we do not have a clean sample of cluster stars.
Despite the possible shortcomings, we have chosen the isochrone fit as a tool
to estimate the reddening of Tr~20 along with its age and distance modulus.
Because the metallicity of Tr~20  seems to be reasonably well-determined,
we chose the \citet{ma08} 
interactive web interface\footnote{http://stev.oapd.inaf.it/cmd}
to generate isochrones to our specifications. A number of
isochrones with convective overshoot of $D_{\rm mix}=0.25H_p$ \citep{gi00}
were built between the ages from 1 through 2 Gyr at steps
of 0.1 Gyr, $Z=0.015$, and for zero interstellar extinction. These
isochrones were used to visually find the best match to various features, 
simultaneously in both $BV$ and $VI$ color-magnitude diagrams. 
These features included the shape of the upper main sequence and
the location of red clump and tentative sub-giant
branch stars (marked by the bold dots in Figs.~1,2 and listed in Table~2).
The optimal fit was achieved with a 1.3 Gyr isochrone.
We note that in a simultaneous adjustment of reddening, age, and distance,
the age is the most sensitive parameter followed by the reddening.
An imperfect knowledge of metallicity [Fe/H] can also affect the reddening
estimate and the color match for RGB stars, although this  plays
a minor role in our study.

To de-redden the observational CMDs, we used a range of possible
$E(B-V)$ values at steps of 0.01 mag. As shown by \citet{fe63},
the reddening is a function of spectral type: 
$\eta$=$E_{B-V}$(Sp~T)$/E_{B-V}$(B0), where $\eta=0.97-0.09(B-V)_{0}$.
Likewise, $E(V-I)/E(B-V)=1.24(1+0.06(B-V)_{0}+0.014E(B-V))$ as
derived by \citet{de78}. Our estimate of reddening for Tr~20, 
extrapolated to a B0 spectral type, is $E(B-V)=0.46$. From the placement
of isochrones in Figs.~1,2 using this reddening estimate, it may
appear that a bit higher reddening would be more appropriate to better match
the main sequence. This is not possible to achieve without
ruining a fit of all the post-main-sequence features. An independent
check of reddening is provided by star 3764 = MG~675.  Its spectroscopic
effective temperature can be converted into the reddening-free colors using
a calibration of normal colors for giants from \citet{be79}. The
corresponding normal colors for 3764 = MG~675 are: $(B-V)_{0}$=1.20
and $(V-I)_{0}$=1.21. Then, the observed reddening for this star is
$E(B-V)$=0.32 and $E(V-I)$=0.43 while the isochrone fit indicates
$E(B-V)$=0.38 and $E(V-I)$=0.54, i.e., we may have over-corrected
the colors of this star by $\sim$20\%. Unable to resolve this
controversy (see also Sect. 3.2), we adopt
$E(B-V)=0.46$ as a compromise value of reddening for Tr~20. 
With this reddening and $A_{v}/E(B-V)$=3.1, the true distance
modulus of Tr~20 is $V_{0}-M_{v}$=12.60 or $d$=3.3~kpc.

The red clump stars provide another check on our estimate of
reddening and distance modulus. As pointed out by \citet{pa98} and
\citet{st98},
the average absolute magnitude of the nearby red clump stars in Cousin's
$I$-bandpass ($M_I=-0.23$) is very stable and nearly independent
of their $V-I$ color. We estimate that our mean magnitude and color
of the red clump stars in Tr~20 is  $V_0=13.50$ and $(V-I)_0=0.92$, respectively
(see Fig.~2). Our distance modulus should be increased by $\sim0.2$ mag
to match the known mean absolute magnitude of the field red clump stars.
A similar effect can reached just by increasing our best value of
$E(B-V)$ by 0.06 mag. These numbers are good indicators of inherent
uncertainties in the distance modulus and reddening (see Table~3).
We note that the stars in the red clump of Tr~20 require
additional radial velocity measurements to identify
binary stars which can easily bias the mean properties of its stars.

We conjecture three possible reasons why our reddening correction,
especially that for the red giants, might be somewhat problematic: 
1) our photometry of red giants is not exactly on the $BVI_C$ system.
Because the observed Landolt standards are virtually reddening-free
(as opposed to the reddened cluster stars) their usage in
the transformation equations can bias our calibrated colors;
2) the de-reddening formulae given by \citet{fe63} may not be as accurate
as needed. These formulae should be checked on all old open clusters
with considerable reddening, e.g., $E(B-V)\ga0.3$;
3) finally, the $T_{\rm eff}$-to-color transformation of theoretical
isochrones for red giants itself can be biased. As pointed out by
\citet{pi04}, in different transformations the colors
of RGB stars may differ by up to 0.2 mag. Which of these possible
reasons is relevant to our study is not possible to disentangle from
our data alone.

\section{Cluster parameters and close analogs of Tr~20}

What we now know about Tr~20 is summarized in Table~3. We intentionally
omitted an estimate of absolute proper motion using the UCAC2 catalog
\citep{za04}. The formal errors of UCAC2 proper motions for
stars fainter than $V$$\sim$14 are reaching 6~mas~yr$^{-1}$, which is
grossly inadequate in obtaining a reliable absolute proper motion of
Tr~20.  The lack of kinematic cluster membership is the main reason for
not attempting to assign any formal error to the age estimate.

\begin{table}
  \caption{Main Parameters of Trumpler~20.}
  \begin{tabular}{@{}lc}
  \hline
   Parameter  & Value\\
 \hline
 Cluster center (J2000.0) & $\alpha$=12$^{\rm h}39\fm52$  
$\delta$=$-60\arcdeg$38$\arcmin$\\
 Galactic coordinates & $\ell$=$301\fdg5$  $b$=$+2\fdg2$\\
 Redddening $E(B-V)$  & 0.46$\pm0.1$ (for B0) \\
 Metallicity [Fe/H] & $-0.11\pm0.13$ \\
 Distance modulus $V_{0}$$-$$M_{v}$ & 12.60$\pm0.2$ \\
 Distance (pc)                  & 3300$\pm300$ \\
 Angular diameter   &     $\sim$16$\arcmin$ \\
 Linear diameter (pc)    &     15 \\
 Cluster members ($M_V$$<$$+4$)   &  $\sim$700\\
 Radial velocity (km~s$^{-1}$)   & $-40.8$ \\
 Age (Gyr)         & 1.3 \\
\hline
\end{tabular}
\end{table}

\begin{figure}
\includegraphics[width=84mm]{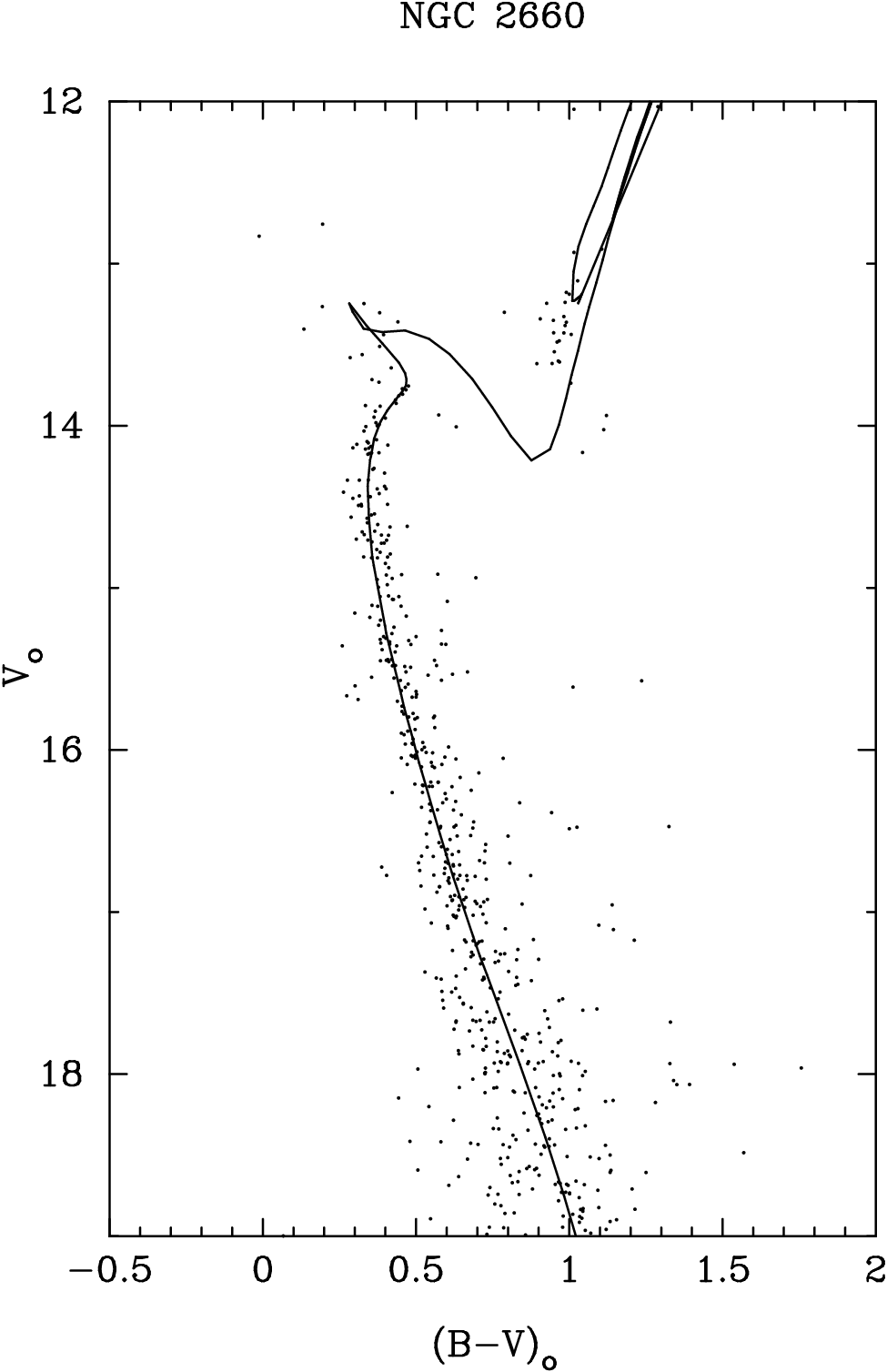}
 \caption{A fit of 1.3-Gyr isochrone to the $BV$ CMD
of NGC~2660. Its true age should be close to 1.3~Gyr despite
a relatively poor match to the red giant clump. The isochrone fit
suggests the reddening $E(B-V)$=0.34 and $V_{0}-M_{V}$=12.15.
}
\end{figure}

\begin{figure}
\includegraphics[width=84mm]{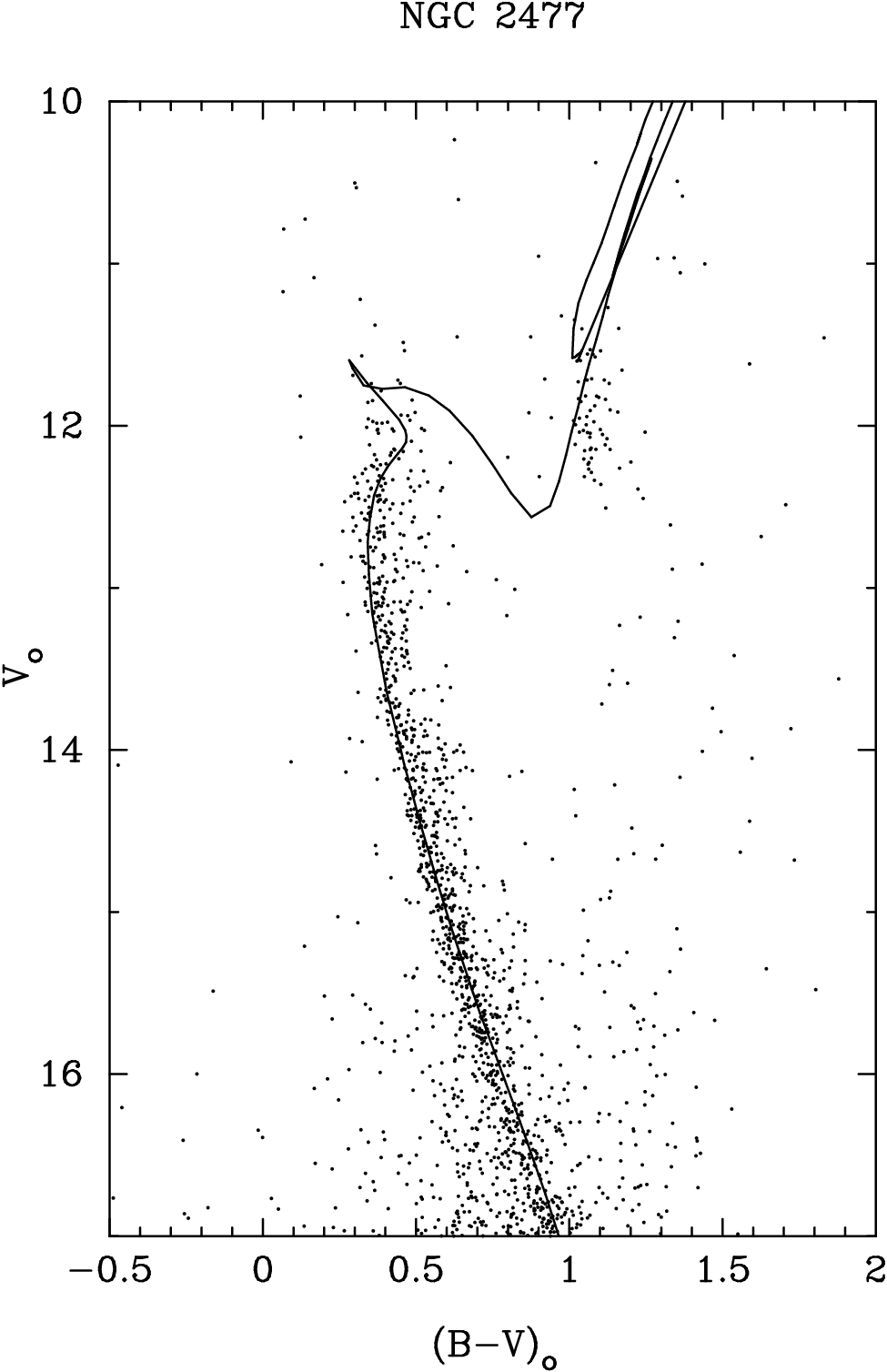}
 \caption{A fit of 1.3-Gyr isochrone to the $BV$ CMD
of NGC~2477. The location of the red-giant clump with respect to this
isochrone clearly favors a younger age.
The isochrone fit suggests the reddening $E(B-V)$=0.16 and $V_{0}-M_{V}$=10.50.
}
\end{figure}

Two rich open clusters NGC~2660 and NGC~2477 -- close analogs of Tr~20 -- were
selected from the WEBDA to illustrate the sensitivity of cluster parameters
upon the isochrone fit. The available CCD $BV$ photometry from \citet{sa99}
for NGC~2660 and from \citet{ka97} for NGC~2477
was de-reddened in the same fashion as for Tr~20. To fit the de-reddened
CMDs (Figs. 3,4), we used a solar-metallicity 1.3~Gyr isochrone with
convective overshoot from \citet{ma08}. 
According to the analysis of high-resolution
spectroscopy, both clusters have essentially a solar metallicity
\citep{br08}. The placement of the red giant clump relative to
the 1.3~Gyr isochrone indicates that the age of NGC~2660 is close to
1.3~Gyr but the age of NGC~2477 is definitely lower than 1.3~Gyr,
in agreement with conclusions from the original studies. However,
we note a need for a lower reddening in both clusters, provided
the metallicity is correct. For a 1.3~Gyr isochrone
near the turnoff of main sequence, an increase of [Fe/H] by 0.1 dex
is equivalent to the decrease of $E(B-V)$ by $\sim$0.035 mag.
Therefore, in deriving the cluster parameters from an isochrone fit,
a prior accurate and consistent metallicity determination is
crucial to this method.

The spatial distribution of stars over the FOV can serve to estimate
the angular extent of Tr~20 and the cumulative number of cluster
members brighter than the limiting magnitude of our study. Figure~5
shows the density of stars as a function of radial distance
(radius $r$) from the cluster center. To construct this distribution,
we used only the stars with reliable $BV$ photometry. It is evident
that the cluster has an angular diameter at least 16$\arcmin$, which
at the nominal distance of Tr~20 translates into 15~pc of linear
diameter. Assuming that the density of field stars down to our
limiting magnitude is about 9 stars per arcmin$^2$, the cumulative
number of cluster members down to $M_V$=$+$4 is $\sim$700. Even this
conservative estimate is indicative of a very rich open cluster.

\begin{figure}
\includegraphics[width=84mm]{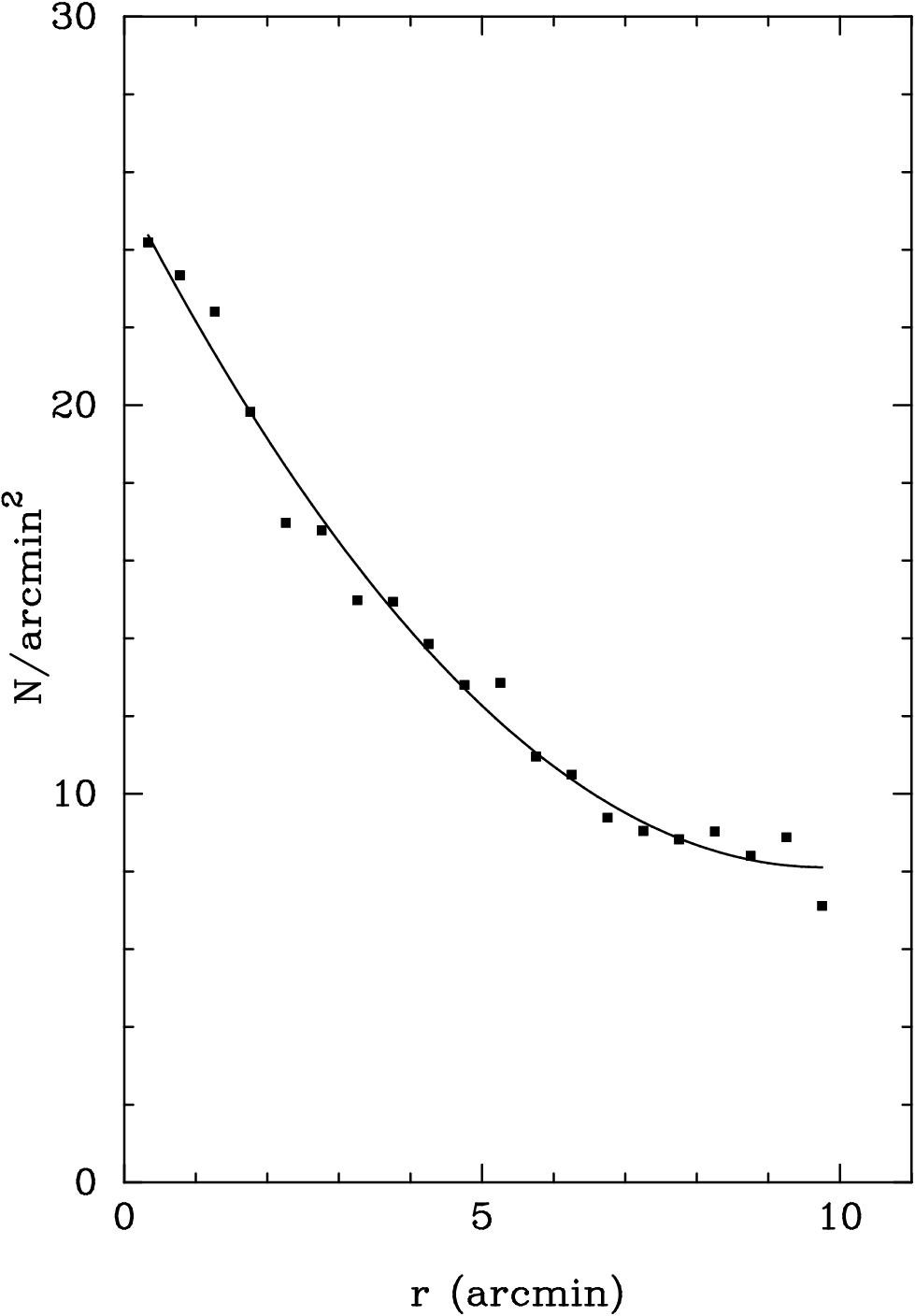}
 \caption{Radial distribution of stars in Tr~20. The spatial density
of stars is plotted as a function or radius. The solid line 
representing a quadratic fit to the points is just for guidance.
It is estimated that the angular diameter of Tr~20 is at least
16$\arcmin$.
}
\end{figure}

\section{Conclusions}

Trumpler 20 has several features which makes it attractive for the
future studies. There is no doubt that Tr~20 is a $\sim$1.3
Gyr old open cluster.
In its age group Tr~20 appears to be the richest Galactic
open clusters. Our isochrone fit by no means can be considered final.
It also carries a burden of correlations between the distance, age,
metallicity, and reddening. Among the parameters derived in this
study, the reddening estimate is clearly the least satisfactory.
We provide potential clues possibly biasing our reddening estimate,
which should help in planning the future photometric observations.
The upper part of CMD requires a rigorous clean-up by
the means of radial velocities to better delineate the upper
main sequence, sub-giant, and giant branches. Although we are
confident that the metallicity of Tr~20 is slightly sub-solar,
it is desirable to obtain high-resolution spectra for the
clump stars and redo the analysis and try to eliminate the apparent
discrepancy between the photometric and spectroscopic $T_{\rm eff}$. 
It is not clear what causes a considerable width of the main sequence,
reaching $\sim$0.3 mag. Is it
differential reddening, large population of binaries, or a combination
of both?  The observed spread of the red clump star colors indicates
that differential reddening in Tr~20 may not exceed $\sim$0.1 mag in
$B-V$.  
There is a hint of the blue straggler population
which photometrically is impossible to disantangle from numerous
field stars. We hope that the future studies will enable refining
the cluster parameters and possibly open up new venues of
research of this rich open cluster.

\section*{Acknowledgments}

We thank the referee, Douglas Gies, for thoughtful and detailed
comments on the manuscript.
I. Platais gratefully acknowledges support from the National Science
Foundation through grant AST 04-06689 to Johns Hopkins University.
J. Fulbright acknowledges support through grants from the W.M. Keck Foundation
and the Gordon and Betty Moore Foundation, to establish  a program of
data-intensive science at the Johns Hopkins University. R. M\'{e}ndez
acknowledges the support by the Chilean Centro de Astrof\'{i}sica
FONDAP (15010003). Support from the Funda\c{c}\~{a}o para Ci\^{e}ncia
e a Tecnologia (Portugal) to P. Figueira in the form of a scholarship
(reference SFRH/BD/21502/2005) is gratefully acknowledged.
This research has made use of the WEBDA database, operated at the Institute for Astronomy of the University of Vienna.

\label{lastpage}

\begin{thebibliography}{}
\bibitem[\protect\citeauthoryear{Baranne et al.}{1996}]{ba96}
Baranne A., Queloz, D., Mayor M., Adrianzyk G., Knispel G., Kohler D.,
Lacroix D., Meunier J.-P., Rimbaud G., Vin A., 1996, A\&AS, 119, 373
\bibitem[\protect\citeauthoryear{Bessell}{1979}]{be79} Bessell, M. S.,
1979, PASP, 91, 589
\bibitem[\protect\citeauthoryear{Bragaglia et al.}{2008}]{br08} Bragaglia, A.,
Sestito, P., Villanova, S., Carretta, E., Randich, S., Tosi, M., 2008,
A\&A, 480, 79
\bibitem[\protect\citeauthoryear{Carraro et al.}{2008}]{ca08}
Carraro G., Villanova S., Demarque P., Moni Bidin C., McSwain M. V.,
2008, MNRAS, 386, 1625
\bibitem[\protect\citeauthoryear{Cousins \& Caldwell}{1985}]{co85}
Cousins A. W. J., Caldwell J. A. R., 1985, Observatory, 105, 134
\bibitem[\protect\citeauthoryear{Dean, Warren \& Cousins}{1978}]{de78}
Dean J. F., Warren P. R., Cousins A. W. J., 1978, MNRAS, 183, 569
\bibitem[\protect\citeauthoryear{Dias et al.}{2004}]{di04}
Dias W. S., L\'{e}pine J. R. D., Alessi B. S., Moitinho A., 2004, Open clusters and Galactic structure, Version 2.0, http://www.astro.iag.usp.br/$^{\sim}$wilton
\bibitem[\protect\citeauthoryear{Fernie}{1963}]{fe63} Fernie J. D., 1963,
AJ, 68, 780
\bibitem[\protect\citeauthoryear{Friel}{1995}]{fr95} Friel, E. D., 1995,
ARA\&A, 33, 381
\bibitem[\protect\citeauthoryear{Fulbright, McWilliam \& Rich}{2006}]{fu06}
Fulbright J. P., McWilliam A., Rich R. M., 2006, ApJ, 636, 821
\bibitem[\protect\citeauthoryear{Girardi et al.}{2000}]{gi00} Girardi L.,
Bressan A., Bertelli G., Chiosi C., 2000, A\&AS, 141, 371
\bibitem[\protect\citeauthoryear{Hogg}{1965}]{ho65} Hogg A. R., 1965,
Mem. Mt. Stromlo Obs., 17, 1
\bibitem[\protect\citeauthoryear{Kassis et al.}{1997}]{ka97} Kassis, M.,
Janes, K. A., Friel, E. D., Phelps, R. L., 1997, AJ, 113, 1723 
\bibitem[\protect\citeauthoryear{Kaufer et al.}{1999}]{ka99}
Kaufer A., Stahl O., Tubbesing S., N{\o}rregaard P., Avila G., Francois P.,
Pasquini L., Pizzella A., 1999, The Messenger, 95, 8
\bibitem[\protect\citeauthoryear{Landolt}{1992}]{la92} Landolt A. U.,
1992, AJ, 104, 340
\bibitem[\protect\citeauthoryear{Marigo et al.}{2008}]{ma08} Marigo P.,
Girardi L., Bressan A., Groenewegen M. A. T., Silva L., Granato G. L.,
2008, A\&A, 482, 883
\bibitem[\protect\citeauthoryear{McSwain \& Gies}{2005}]{mc05} 
McSwain M. V., Gies D. R., 2005, ApJS, 161, 118
\bibitem[\protect\citeauthoryear{Paczy\'{n}ski \& Stanek}{1998}]{pa98}
Paczy\'{n}sky B., Stanek K. Z., 1998, ApJ, 494, L219
\bibitem[\protect\citeauthoryear{Pietrinferni et al.}{2004}]{pi04}
Pietrinferni A., Cassisi S., Salaris M., Castelli F., 2004, ApJ, 612, 168
\bibitem[\protect\citeauthoryear{Platais et al.}{2007}]{pl07} Platais I.,
Melo C., Mermilliod J.-C., Kozhurina-Platais V., Fulbright J. P.,
M\'{e}ndez R. A., Altmann M., Sperauskas J., 2007, A\&A, 461, 509
\bibitem[\protect\citeauthoryear{Queloz}{1995}]{qu95} Queloz D., 1995,
in A. G. D. Philips, K. A. Janes \& A. R. Upgren, eds, Proc. IAU Symp. 167,
New Developments in Array Technology and Applications,
Kluwer, Dordrecht, p. 221
\bibitem[\protect\citeauthoryear{Ram\'{i}rez \& Mel\'{e}ndez}{2005}]{ra05}
Ram\'{i}rez I., Mel\'{e}ndez J., 2005, ApJ, 626, 465
\bibitem[\protect\citeauthoryear{Sandrelli et al.}{1999}]{sa99}  Sandrelli S.,
Bragaglia A., Tosi M., Marconi G., 1999, MNRAS, 309, 739
\bibitem[\protect\citeauthoryear{Stanek \& Garnavich}{1998}]{st98} Stanek K. Z., Garnavich P. M., 1998, ApJ, 503, L131
\bibitem[\protect\citeauthoryear{Taylor}{2007}]{ta07} Taylor B. J., 2007,
AJ, 133, 370
\bibitem[\protect\citeauthoryear{Trumpler}{1930}]{tr30}  Trumpler R. J.,
1930, Lick. Obs. Bull., 14, 154
\bibitem[\protect\citeauthoryear{Zacharias et al.}{2004}]{za04} 
Zacharias N., Urban S. E., Zacharias M. I., Wycoff G. L., Hall D. G.,
Monet D. G., Rafferty T. J., 2004, AJ, 127, 3043

\end{thebibliography}
\end{document}